\documentclass[
aps,
prb,
onecolumn,
]{revtex4-2}

\usepackage{amsmath,amssymb,amsthm}
\usepackage{physics}
\usepackage{graphicx}
\usepackage{hyperref}
\usepackage{xcolor}
\usepackage{booktabs}
\usepackage{bbm}

\newtheorem{theorem}{Theorem}
\newtheorem{proposition}[theorem]{Proposition}
\newtheorem{corollary}[theorem]{Corollary}

\newcommand{\Hrot}{\tilde{H}}
\hypersetup{colorlinks=true, linkcolor=blue, citecolor=blue, urlcolor=blue}

\begin{document}

\title{Implications of the PXP ground-state sign structure for neural-network variational Monte Carlo}


\author{Jun Sur Richard Park}
\email[Correspondence to: ]{junsurpark@korea.ac.kr}
\affiliation{Division of Big Data Science, Korea University, Sejong, 30019, Republic of Korea }

\author{Kwangmin Yu}
\email{kyu@bnl.gov}
\affiliation{Computational Science Department, Brookhaven National Laboratory, Upton, New York 11973, USA}

\date{}

\begin{abstract}
Positive-definite neural-network quantum states cannot directly represent the ground state of the PXP Hamiltonian, since the ground-state wavefunction carries a nontrivial relative-sign structure. Building on the chiral symmetry of the PXP Hamiltonian, we show that this sign structure is fixed exactly by the parity of the total up-spin number: a diagonal unitary transformation that assigns a sign to each spin configuration according to this parity maps the PXP Hamiltonian to its negative, rendering the transformed Hamiltonian stoquastic on the Rydberg-blockaded Hilbert space. Since the configuration graph of allowed configurations is connected, the Perron--Frobenius theorem guarantees a unique, strictly positive ground state of the transformed Hamiltonian, yielding an exact sign rule for the ground state of the original Hamiltonian in terms of the parity of the up-spin number. Variational Monte Carlo performed directly on the transformed Hamiltonian with a positive-definite ansatz therefore requires no architecture-specific modification to represent the true ground state. We evaluate the performance of this formulation using five representative neural-network quantum state architectures, SineKAN, reflection-symmetrized SineKAN, RBM, MLP, and LSTM, for different chain lengths.
\end{abstract}

\maketitle

\section{Introduction}

The PXP model describes a chain of spin-$1/2$ particles subject to the Rydberg blockade constraint and has emerged as a paradigmatic model for studying quantum many-body scars \cite{turner2018weak}. Unlike generic interacting quantum systems that obey the eigenstate thermalization hypothesis (ETH), the PXP model hosts special nonthermal eigenstates responsible for long-lived coherent oscillations following quenches from certain product states \cite{bernien2017probing,ho2019periodic}. Consequently, much of the literature has focused on nonequilibrium dynamics, quantum revivals, and scarred eigenstates. In contrast, the use of neural-network quantum states to approximate the PXP ground state has received comparatively less attention.

Neural-network quantum states (NQS) provide a flexible variational framework for quantum many-body problems. Since the pioneering work of Carleo and Troyer \cite{carleo2017solving}, variational Monte Carlo (VMC) with NQS has achieved high accuracy for a broad range of lattice models. An important consideration in applying such methods is the sign or phase structure of the target wavefunction. In particular, an ansatz constrained to have strictly positive real amplitudes cannot directly represent a target state with nontrivial relative signs unless the appropriate sign structure is incorporated explicitly or removed by a basis transformation.

A classical example is the Marshall sign rule \cite{marshall1955antiferromagnetism}, which identifies the ground-state sign structure of the antiferromagnetic Heisenberg model on bipartite lattices and provides a basis in which the relevant Hamiltonian is stoquastic. The PXP model has a different structure: unlike the Heisenberg model, it does not conserve total magnetization, since its dynamics are generated by kinetically allowed single-spin flips, so the relevant grading is the parity of the total excitation number rather than a sublattice decomposition. The PXP Hamiltonian is furthermore known to possess a chiral symmetry associated with this parity \cite{buijsman2022number}. Here, rather than treating this symmetry as a new property of the PXP model, we focus on its implications for the ground-state sign structure and for positive-amplitude variational calculations.

Specifically, we consider the diagonal parity transformation
\(
D|\sigma\rangle
=
(-1)^{N_\uparrow(\sigma)}
|\sigma\rangle.
\)
Because every allowed PXP transition changes $N_\uparrow$ by one, this transformation gives
\(
\tilde{H}=DHD=-H.
\)
On the Rydberg-blockaded Hilbert space, the transformed Hamiltonian is stoquastic and its configuration graph is connected. The Perron--Frobenius theorem\cite{horn2012matrix} then guarantees that the ground state $|\Phi_0\rangle$ of $\tilde{H}$ can be chosen to have strictly positive amplitudes on every allowed configuration. Since
\(
|\Phi_0\rangle=D|\psi_0\rangle,
\)
where $|\psi_0\rangle$ is the ground state of the original Hamiltonian, it follows that
\(
\psi_0(\sigma)
=
(-1)^{N_\uparrow(\sigma)}
\Phi_0(\sigma),
\,
\Phi_0(\sigma)>0.
\)
Thus the parity structure associated with the chiral symmetry directly determines the relative signs of the PXP ground-state amplitudes.

This observation has a direct consequence for VMC. A strictly positive variational ansatz applied directly to the original Hamiltonian cannot reproduce the required relative signs unless the factor $(-1)^{N_\uparrow(\sigma)}$ is included explicitly. Equivalently, one may transform the problem and perform VMC directly on
\(
\tilde{H}=-H
\)
using a positive-definite real-valued ansatz. The resulting variational state approximates the positive ground state $|\Phi_0\rangle$, while the corresponding approximation to the original PXP ground state is recovered by applying $D$. This rotated formulation requires no architecture-specific modification, since any positive-amplitude ansatz can be applied directly to $\tilde{H}$ without additional sign-structure engineering.

To demonstrate this formulation numerically, we apply five representative NQS architectures, RBM, MLP, LSTM, SineKAN, and its reflection-symmetrized variant rSineKAN \cite{shamim2025probing}, to the transformed Hamiltonian for chain lengths up to $L=64$. These architectures provide different parameterizations of positive-amplitude variational states and allow us to examine the applicability of the transformed formulation across different classes of NQS. We evaluate their variational energies and energy variances against exact diagonalization for $L\le20$ and against DMRG for larger $L$, and, where exact diagonalization is available, their ground-state fidelities and correlation functions. The purpose of these experiments is to demonstrate the practical use of the sign transformation rather than to establish a ranking among neural-network architectures.

\section{Ground-State Sign Structure of the PXP model}

In this section we show that a diagonal parity transformation maps the PXP Hamiltonian to its negative, which is stoquastic. Together with the connectivity of the configuration graph, the Perron--Frobenius theorem then fixes the sign of every ground-state amplitude of the original Hamiltonian exactly.

\subsection{Hamiltonian and constrained Hilbert space}

We consider a chain of $L$ spin-$1/2$ particles with open boundary conditions. The full spin Hilbert space is
\(
\mathcal{H}_{\mathrm{full}}
=
(\mathbb{C}^2)^{\otimes L},
\)
spanned by computational-basis states
\(
|\sigma\rangle
=
|\sigma_1\sigma_2\cdots\sigma_L\rangle,
\ \,
\sigma_j\in\{\uparrow,\downarrow\}.
\)

The PXP model is considered on the Rydberg-blockaded subspace
\[
\mathcal{H}_{\mathrm{PXP}}
=
\operatorname{span}
\left\{
|\sigma\rangle\in\mathcal{H}_{\mathrm{full}}:
n_jn_{j+1}=0
\quad\text{for all }j
\right\},
\]
where
\(
n_j=\frac{1+\sigma_j^z}{2}.
\)
Thus configurations containing two adjacent up-spins are excluded. Throughout this work, unless otherwise stated, the Hamiltonian and all variational states are understood to be restricted to $\mathcal{H}_{\mathrm{PXP}}$.

The PXP Hamiltonian is
\(
H
=
\sum_{j=1}^{L}
P_{j-1}^{\downarrow}
X_j
P_{j+1}^{\downarrow},
\)
where
\(
X_j=\sigma_j^x,
\,
P_j^\downarrow
=
\frac{\mathbbm{1}-\sigma_j^z}{2},
\)
with
\(
P_0^\downarrow=P_{L+1}^\downarrow=\mathbbm{1}
\)
for the boundaries.
In the computational basis of $\mathcal{H}_{\mathrm{PXP}}$, $H$ is purely off-diagonal, and every nonzero matrix element is $+1$. Each nonzero matrix element corresponds to an allowed single-spin flip. Therefore,
\[
\langle\sigma'|H|\sigma\rangle\neq0
\quad\Longrightarrow\quad
N_\uparrow(\sigma')
=
N_\uparrow(\sigma)\pm1,
\]
where
\(
N_\uparrow(\sigma)
=
\sum_{j=1}^{L}
\frac{1+\sigma_j^z}{2}.
\)

The configuration graph is therefore bipartite with respect to the parity of $N_\uparrow$: every Hamiltonian edge connects an even-$N_\uparrow$ configuration to an odd-$N_\uparrow$ configuration.

\subsection{Parity transformation and stoquastic Hamiltonian}

Define the diagonal operator
\begin{equation}
D
=
\sum_{\sigma\in\mathcal{H}_{\mathrm{PXP}}}
s_\sigma
|\sigma\rangle\langle\sigma|,
\qquad
s_\sigma=(-1)^{N_\uparrow(\sigma)}.
\end{equation}
Since $s_\sigma=\pm1$, the operator $D$ is Hermitian and unitary,
\(
D^\dagger=D,
\,
D^2=\mathbbm{1}.
\)

\begin{proposition}\label{prop:stoquastic}
The transformed Hamiltonian
\(
\Hrot=DHD
\)
satisfies
\(
\Hrot=-H.
\)
Consequently, $\Hrot$ is stoquastic in the computational basis of $\mathcal{H}_{\mathrm{PXP}}$.
\end{proposition}

\begin{proof}
Consider two configurations with
\(
\langle\sigma'|H|\sigma\rangle\neq0.
\)
Since $H$ connects configurations differing by one allowed spin flip,
\(
N_\uparrow(\sigma')
=
N_\uparrow(\sigma)\pm1.
\)
Therefore
\(
s_{\sigma'}s_\sigma
=
(-1)^{N_\uparrow(\sigma')+N_\uparrow(\sigma)}
=
-1.
\)
It follows that
\[
\langle\sigma'|DHD|\sigma\rangle
=
s_{\sigma'}
\langle\sigma'|H|\sigma\rangle
s_\sigma
=
-\langle\sigma'|H|\sigma\rangle.
\]
Every matrix element of $H$ that does not vanish therefore changes sign under $D$, while vanishing matrix elements are unaffected. Hence $DHD=-H$. 
Because every nonzero matrix element of $H$ is positive, every nonzero matrix element of $\Hrot=-H$ is negative, and since $H$ has no diagonal elements, 
these are all off-diagonal. Hence $\Hrot$ is stoquastic.
\end{proof}

\subsection{Connectivity and Perron--Frobenius positivity}

Stoquasticity alone guarantees that the ground state can be chosen nonnegative within each irreducible sector. To obtain strict positivity over the entire constrained basis, we also use the connectivity of the PXP configuration graph\cite{jeevanesan2023quantum}. For completeness, we include a short proof.

\begin{proposition}\label{prop:connected}
The configuration graph generated by $H$ on $\mathcal{H}_{\mathrm{PXP}}$ is connected.
\end{proposition}

\begin{proof}
Take any allowed configuration $|\sigma\rangle\in\mathcal{H}_{\mathrm{PXP}}$ containing at least one up-spin. In an allowed configuration, every neighbor of an up-spin is a down-spin, so the projectors in the corresponding term of $H$ both equal one and the term flips this spin down. This gives a configuration $|\sigma'\rangle$ with $\langle\sigma'|H|\sigma\rangle\neq0$. Flipping an up-spin down never creates a pair of adjacent up-spins, so $|\sigma'\rangle$ is again allowed, and it has one fewer up-spin than $|\sigma\rangle$. Repeating this at most $N_\uparrow(\sigma)$ times connects $|\sigma\rangle$ to the all-down state
\(
|\downarrow\downarrow\cdots\downarrow\rangle.
\)
Since $H$ is symmetric, every edge of the graph can be traversed in both directions. Any two allowed configurations are therefore connected through the all-down state, and the configuration graph is connected.
\end{proof}

\begin{theorem}\label{thm:pf}
The ground state of $\Hrot=-H$ on $\mathcal{H}_{\mathrm{PXP}}$ is nondegenerate and can be chosen to have strictly positive amplitudes in every allowed computational-basis configuration.
\end{theorem}
\begin{proof}
Since $\mathcal{H}_{\mathrm{PXP}}$ is finite-dimensional, we may represent $H$ as a finite
matrix in the computational basis. By construction, $H$ is nonnegative: its diagonal
entries vanish identically, since $H$ contains no diagonal terms, and its nonzero
off-diagonal entries equal $+1$. By Proposition~\ref{prop:connected}, the configuration
graph associated with $H$ is connected; since $H$ is real and symmetric, this
connectivity is equivalent to irreducibility of $H$ in the sense required by the
Perron--Frobenius theorem\cite{horn2012matrix}.

The Perron--Frobenius theorem therefore guarantees that the largest eigenvalue of $H$ is
nondegenerate and that a corresponding eigenvector can be chosen strictly positive on
every configuration $\sigma\in\mathcal{H}_{\mathrm{PXP}}$.

It remains to identify this eigenvector with the ground state of $\Hrot$. By
Proposition~\ref{prop:stoquastic}, $\Hrot=-H$, so if $H|\phi\rangle=\mu|\phi\rangle$, then
\(
\Hrot|\phi\rangle=-H|\phi\rangle=-\mu|\phi\rangle,
\)
so $H$ and $\Hrot$ share the same eigenvectors, with the eigenvalue of $\Hrot$ given by
$-\mu$. Thus the eigenvector realizing the \emph{largest} eigenvalue of $H$ is exactly the
eigenvector realizing the \emph{smallest} eigenvalue of $\Hrot$.

Combining the two statements, the ground state of $\Hrot$ is unique and strictly positive
in the computational basis of $\mathcal{H}_{\mathrm{PXP}}$.
\end{proof}

Let this ground state be denoted by $|\Phi_0\rangle$:
\begin{equation}
\Hrot|\Phi_0\rangle
=
E_0|\Phi_0\rangle,
\qquad
\Phi_0(\sigma)>0
\quad
\text{for all }
\sigma\in\mathcal{H}_{\mathrm{PXP}}.
\label{eq:phi_ground}
\end{equation}

\subsection{Ground-state sign structure of the original Hamiltonian}

Let $|\psi_0\rangle$ denote the ground state of the original PXP Hamiltonian:
\(
H|\psi_0\rangle
=
E_0|\psi_0\rangle.
\)
Because
\(
\Hrot=DHD
\)
is unitarily equivalent to $H$, the two Hamiltonians have identical spectra. Moreover,
\(
\Hrot D|\psi_0\rangle
=
DHD D|\psi_0\rangle
=
DH|\psi_0\rangle
=
E_0D|\psi_0\rangle.
\)
Thus $|\Phi_0\rangle=D|\psi_0\rangle$ is an eigenvector of $\Hrot$ with eigenvalue $E_0$, which is the lowest eigenvalue of $\Hrot$ since the two spectra coincide due to unitary equivalence. 
It is therefore the ground state of $\Hrot$, with the same ground-state energy.

Since $D^2=\mathbbm{1}$, we have
\(
|\psi_0\rangle
=
D|\Phi_0\rangle.
\)
Therefore
\(
\psi_0(\sigma)
=
(-1)^{N_\uparrow(\sigma)}
\Phi_0(\sigma).
\)
Theorem~\ref{thm:pf} gives
\(
\Phi_0(\sigma)>0
\)
for every allowed configuration. Hence:

\begin{corollary}\label{cor:sign}
The exact ground-state sign structure of the PXP Hamiltonian on $\mathcal{H}_{\mathrm{PXP}}$ is
\[
\boxed{
\mathrm{sgn}\,\psi_0(\sigma)
=
(-1)^{N_\uparrow(\sigma)}
}
\]
for every allowed configuration $\sigma$, up to an overall global sign.
\end{corollary}

Thus amplitudes with even $N_\uparrow$ have one sign and amplitudes with odd $N_\uparrow$ have the opposite sign. A strictly positive ansatz applied directly to $H$ in the original computational basis cannot represent this relative-sign structure unless the known factor $(-1)^{N_\uparrow(\sigma)}$ is incorporated explicitly.

\subsection{Connection to chiral symmetry}

The transformation above is directly related to the chiral symmetry of the PXP model. The Hamiltonian anticommutes with the parity operator $C=(-1)^L\prod_{j=1}^{L}\sigma_j^z$~\cite{buijsman2022number}, which acts on a computational-basis state as $C|\sigma\rangle=(-1)^{N_\uparrow(\sigma)}|\sigma\rangle$ and therefore coincides with $D$. Indeed, $\prod_j\sigma_j^z$ gives $+1$ for each up spin and $-1$ for each down spin, so that $\prod_j\sigma_j^z|\sigma\rangle=(-1)^{L-N_\uparrow(\sigma)}|\sigma\rangle$, and the prefactor $(-1)^L$ cancels the $L$-dependent sign for every chain length.
The identity $DHD=-H$ of Proposition~\ref{prop:stoquastic} is thus the chiral symmetry of the model expressed in terms of the parity of the total up-spin number.

This identity connects the ground states of $H$ and $-H$. Let $|\Phi_0\rangle$ be the ground state of $-H$, with $(-H)|\Phi_0\rangle=E_0|\Phi_0\rangle$. Since $H=D(-H)D$ and $D^2=\mathbbm{1}$,
\(
H\,D|\Phi_0\rangle
=
D(-H)|\Phi_0\rangle
=
E_0\,D|\Phi_0\rangle,
\)
so $D|\Phi_0\rangle$ is an eigenvector of $H$ with eigenvalue $E_0$. Because $H$ and $-H$ are unitarily equivalent and have the same spectrum, $E_0$ is also the lowest eigenvalue of $H$, and $D|\Phi_0\rangle$ is the ground state of $H$. Hence
\begin{equation}
\boxed{
|\psi_0\rangle=D|\Phi_0\rangle,
\qquad
E_0(H)=E_0(-H).
}
\label{eq:ground_relation}
\end{equation}
Note that $|\Phi_0\rangle$ itself is not the ground state of $H$ but its highest-energy eigenstate, so the factor $D$ cannot be omitted.

Since $DHD=-H$, VMC can be performed directly on $-H$ using a positive-definite ansatz to approximate $|\Phi_0\rangle$, without constructing $DHD$ explicitly. The corresponding approximation to the ground state of the original Hamiltonian is obtained by applying $D$.

\subsection{Gauge invariance of fidelity and diagonal observables}

\begin{proposition}[Gauge invariance of fidelity and diagonal observables]
\label{prop:gauge_invariance}
Let $|\Phi_0\rangle$ be the ground state of $\Hrot$ and let $|\Phi_{\rm VMC}\rangle$ be a variational approximation to $|\Phi_0\rangle$ for $\Hrot$. Define $|\psi_0\rangle=D|\Phi_0\rangle$, which is the ground state of $H$, and $|\psi_{\rm VMC}\rangle=D|\Phi_{\rm VMC}\rangle$. Then
\[
\frac{
|\langle\Phi_0|\Phi_{\rm VMC}\rangle|^2
}{
\langle\Phi_0|\Phi_0\rangle
\langle\Phi_{\rm VMC}|\Phi_{\rm VMC}\rangle
}
=
\frac{
|\langle\psi_0|\psi_{\rm VMC}\rangle|^2
}{
\langle\psi_0|\psi_0\rangle
\langle\psi_{\rm VMC}|\psi_{\rm VMC}\rangle
}.
\]
The expectation value of any observable diagonal in the computational basis is also invariant under the transformation.
\end{proposition}

\begin{proof}
Since $D^\dagger D=\mathbbm{1}$, we have
\(
\langle\psi_0|\psi_{\rm VMC}\rangle
=
\langle\Phi_0|
D^\dagger D
|\Phi_{\rm VMC}\rangle
=
\langle\Phi_0|\Phi_{\rm VMC}\rangle.
\)
The normalization factors are unchanged for the same reason.

For a diagonal observable
\(
\hat{O}
=
\sum_\sigma
O(\sigma)
|\sigma\rangle\langle\sigma|,
\)
the transformation only multiplies each amplitude by $s_\sigma=\pm1$, and therefore
\(
|\psi(\sigma)|^2
=
|\Phi(\sigma)|^2
\)
for every $\sigma$. Since
\(
\langle\hat{O}\rangle
=
\sum_\sigma O(\sigma)|\psi(\sigma)|^2\big/\sum_\sigma|\psi(\sigma)|^2,
\)
diagonal expectation values are unchanged.
\end{proof}

This result justifies computing the fidelities and diagonal correlation functions reported below directly using the ground state of $\Hrot=-H$, without explicitly transforming the states back to the original basis.

\section{Implications for Variational Monte Carlo}
\label{sec:vmc_implications}

The preceding results clarify the role of the sign transformation in variational calculations.

\paragraph{Positive real ansatz.}
The exact ground state $|\Phi_0\rangle$ of $\Hrot=-H$ is strictly positive on every allowed configuration. Therefore, there is no sign-structure obstruction to approximating $|\Phi_0\rangle$ with a positive-definite real-valued variational state. A complex phase is not required by the exact target state, although the expressive and optimization properties of a particular finite neural-network architecture remain architecture dependent.

Equivalently, one may work directly with the original Hamiltonian $H$ using an ansatz of the form
\(
\psi_\theta(\sigma)
=
(-1)^{N_\uparrow(\sigma)}
\exp[f_\theta(\sigma)].
\)
Working with $\Hrot=-H$ instead absorbs this known sign structure into the Hamiltonian and allows the variational model itself to remain positive.

\paragraph{Local-energy sign structure.}
For a positive ansatz $\psi(\sigma)>0$, the local energy of $\Hrot$ is
\begin{equation}
E_{\rm loc}(\sigma)
=
\sum_{\sigma'}
\langle\sigma'|\Hrot|\sigma\rangle
\frac{\psi(\sigma')}{\psi(\sigma)}.
\end{equation}
Because the nonzero matrix elements of $\Hrot$ are nonpositive and $\psi(\sigma')/\psi(\sigma)>0$, all contributions to
$E_{\rm loc}(\sigma)$ have the same sign, so no sign cancellation occurs among them. In contrast, when optimizing the original
Hamiltonian $H$ without prior knowledge of its ground-state sign structure, the variational wavefunction may assign different relative
signs to connected configurations, so that the ratios $\psi(\sigma')/\psi(\sigma)$ can take both signs and contributions of
opposite signs can partially cancel. When such cancellations are substantial, the local energy depends on a balance between positive
and negative contributions and can therefore be sensitive to errors in their relative magnitudes. For $\Hrot$ with a positive ansatz, this
source of sensitivity is absent because all contributions enter with a definite nonpositive sign.

\subsection{Variational ans\"atze}

To illustrate the transformed formulation with different classes of NQS, we consider five representative architectures: SineKAN, a reflection-symmetrized variant of SineKAN, RBM, MLP, and LSTM. All models output a real-valued log-amplitude
\(
f_\theta(\sigma)
=
\log\psi_\theta(\sigma)
\in\mathbb{R},
\)
so that
\(
\psi_\theta(\sigma)
=
\exp[f_\theta(\sigma)]
>0.
\)
By Theorem~\ref{thm:pf}, this positivity constraint is compatible with the exact ground state of $\Hrot$.

\paragraph{SineKAN.}
SineKAN \cite{shamim2025probing} is a neural quantum-state architecture based on Kolmogorov--Arnold Networks (KANs) \cite{liu2025kan} with learnable sinusoidal activations. A SineKAN is built by stacking layers, each of which computes
\begin{equation}
y_m
=
\sum_{l,n}
A_{mnl}
\sin\!\left(
\omega_n x_l
+
\frac{\pi n}{N}
+
\frac{\pi l}{L}
+
\delta_{nl}
\right)
+
b_m,
\label{eq:sinekan_layer}
\end{equation}
where $x_l$ are the inputs, $A_{mnl}$ are learnable amplitudes, $\omega_n$ are learnable frequencies, $N$ is the grid size, $\delta_{nl}$ is a small random perturbation, and $b_m$ is a bias. The output of the final layer is the log-amplitude $S(\sigma)$ of the SineKAN ansatz.

\paragraph{Reflection-symmetrized SineKAN (rSineKAN).}
The open-chain PXP Hamiltonian is invariant under reflection about the chain center. Reflection leaves $N_\uparrow$ unchanged and therefore commutes with $D$, so $\Hrot$ is reflection invariant as well. We exploit this symmetry by constructing a reflection-symmetrized variant of SineKAN, following Ref.~\cite{shamim2025probing}. Denoting the site-reversal operator by $R$, defined by $(R\sigma)_j=\sigma_{L+1-j}$, the rSineKAN log-amplitude is
\begin{equation}
f_\theta^{\rm rSineKAN}(\sigma)
=
S(\sigma)
+
S(R\sigma),
\label{eq:rsinekan}
\end{equation}
where both terms are evaluated using the same underlying SineKAN weights. Since $R^2=\mathbbm{1}$, this construction gives $f_\theta^{\rm rSineKAN}(\sigma)=f_\theta^{\rm rSineKAN}(R\sigma)$ for every $\sigma$, so the resulting amplitude is manifestly invariant under reflection. Because the two terms in Eq.~\eqref{eq:rsinekan} share the same weights, the number of parameters of rSineKAN is set entirely by the width of the underlying SineKAN network, independent of the number of reflected evaluations.

\paragraph{Other NQS architectures.}
In addition to the two SineKAN variants, we consider a restricted Boltzmann machine (RBM) \cite{carleo2017solving}, a multi-layer perceptron (MLP), and an autoregressive LSTM \cite{graves2012long,hibat2020recurrent}. All five architectures provide structurally different parameterizations of positive-amplitude wavefunctions. RBM, MLP, and the SineKAN variants use MetropolisLocal sampling, whereas the LSTM uses autoregressive direct sampling. Because the architectures and sampling procedures differ, the calculations below are intended as demonstrations of the transformed positive-amplitude formulation rather than as a controlled assessment of architecture superiority.

\section{Numerical Results}
\label{sec:numerics}

We numerically investigate the positive-amplitude formulation by applying VMC to the transformed PXP Hamiltonian
\(
\Hrot=-H
\)
on open chains. Two sets of chain lengths are considered. The first, $L\in\{8,12,16,20\}$, is small enough for exact diagonalization, so that the energy error, the energy variance, the fidelity, and the spin-spin correlation function can all be evaluated against the exact ground state. The second, $L\in\{32,48,64\}$, extends the energy error and variance comparison to larger systems. In this case, exact diagonalization is no longer feasible and the reference energy is obtained from the density-matrix renormalization group (DMRG). The training and reference-calculation settings are given in Appendix~\ref{app:hyperparams}.

\subsection{Exact-diagonalization range}

For each $L\le20$ we diagonalize $\Hrot=-H$ using Lanczos to obtain its exact ground state $|\Phi_0\rangle$, which we denote by $\psi_{\rm ED}$ in the numerical results. As established in Eq.~\eqref{eq:ground_relation}, the ground state of the original Hamiltonian is $|\psi_0\rangle=D|\Phi_0\rangle$. The two states are ground states of different Hamiltonians, $H$ and $\Hrot$, but their ground-state energies are identical,
\(
E_0(\Hrot)=E_0(H).
\)
By Proposition~\ref{prop:gauge_invariance}, the fidelity between exact and variational states is unchanged when both states are transformed by $D$, and observables diagonal in the computational basis, including the spin-spin correlations considered below, are likewise invariant. We may therefore evaluate these quantities directly with the ground state of $\Hrot$.

All five architectures introduced above are parameterized as positive-amplitude states and trained with the Adam optimizer. The learning rate and architectural hyperparameters (grid size, hidden-layer width, RBM $\alpha$, and LSTM feature and layer count) differ across models and chain lengths, and the complete settings are given in Appendix~\ref{app:hyperparams}. RBM, SineKAN, rSineKAN, and MLP use MetropolisLocal sampling, whereas the LSTM uses autoregressive direct sampling.

Table~\ref{tab:pxp_summary} reports results at $L=20$. We list the number of variational parameters, the relative energy error
\(
\epsilon
=
|E_{\rm VMC}-E_{\rm ED}|/|E_{\rm ED}|,
\)
the energy variance
\(
\sigma^2
=
\langle\Hrot^2\rangle-\langle\Hrot\rangle^2,
\)
and the ground-state fidelity
\(
F
=
|\langle\psi_{\rm ED}|\psi_{\rm VMC}\rangle|^2
/
(\langle\psi_{\rm ED}|\psi_{\rm ED}\rangle
\langle\psi_{\rm VMC}|\psi_{\rm VMC}\rangle)
\).

\begin{table}[h]
\centering
\caption{VMC results at $L=20$ for the transformed Hamiltonian $\Hrot=-H$. The reference is the exact ground state obtained by Lanczos diagonalization.}
\label{tab:pxp_summary}
\begin{tabular}{lrrrr}
\toprule
Model & Parameters & $\epsilon$ & $\sigma^2$ & $F$ \\
\midrule
LSTM     & 21,272 & $1.11\times10^{-2}$ & $2.84\times10^{-1}$ & 0.932030 \\
MLP      & 22,821 & $1.00\times10^{-4}$ & $6.72\times10^{-4}$ & 0.999975 \\
RBM      & 53,780 & $1.69\times10^{-2}$ & $4.54\times10^{-3}$ & 0.866534 \\
SineKAN  & 20,121 & $6.95\times10^{-5}$ & $1.68\times10^{-3}$ & 0.999841 \\
rSineKAN & 20,121 & $4.69\times10^{-6}$ & $1.06\times10^{-4}$ & 0.999996 \\
\bottomrule
\end{tabular}
\end{table}

MLP, SineKAN, and rSineKAN reach fidelities above $0.9998$ and relative errors of order $10^{-4}$ or below, while LSTM and RBM are less accurate, with errors of order $10^{-2}$ and fidelities of $0.93$ and $0.87$, respectively. The Perron--Frobenius result guarantees that none of the positive-amplitude architectures faces an intrinsic incompatibility with the sign structure of the exact ground state of $\Hrot$. The differences among the models therefore cannot be attributed to the sign structure, and are more likely due to finite-model expressivity, parameterization, sampling, and optimization.

\begin{figure}[h]
\centering
\includegraphics[width=0.6\textwidth]{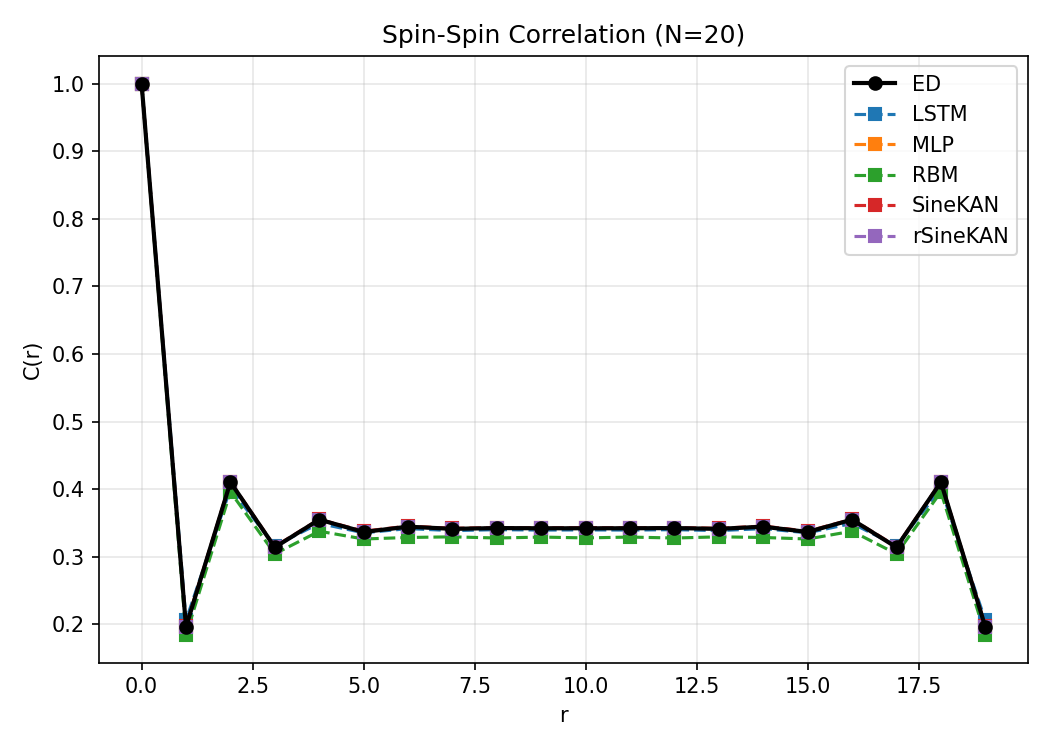}
\caption{Spin-spin correlation $C(r)$ at $L=20$ for the variational ground state of $\Hrot=-H$ and exact diagonalization.}
\label{fig:pxp_corr}
\end{figure}

Figure~\ref{fig:pxp_corr} compares the spin-spin correlation
\(
C(r)
=
\frac{1}{L}
\sum_\ell
\langle
\sigma_\ell^z
\sigma_{\ell+r}^z
\rangle
\)
(with site indices taken modulo $L$) obtained from the variational calculation with the exact-diagonalization result at $L=20$. The structure is reproduced by all five architectures. The curves of MLP, SineKAN, and rSineKAN are indistinguishable from the exact result on the scale of the figure, while the RBM curve lies slightly below it over the plateau, and the LSTM curve deviates slightly at the shortest distances, in line with the lower fidelities of these two models in Table~\ref{tab:pxp_summary}. The variational states therefore describe not only the ground-state energy but also a diagonal observable of the transformed ground state. Because this observable is invariant under $D$, the same correlation function is obtained for the corresponding state of the original Hamiltonian.

\begin{figure}[h]
\centering
\includegraphics[width=0.6\textwidth]{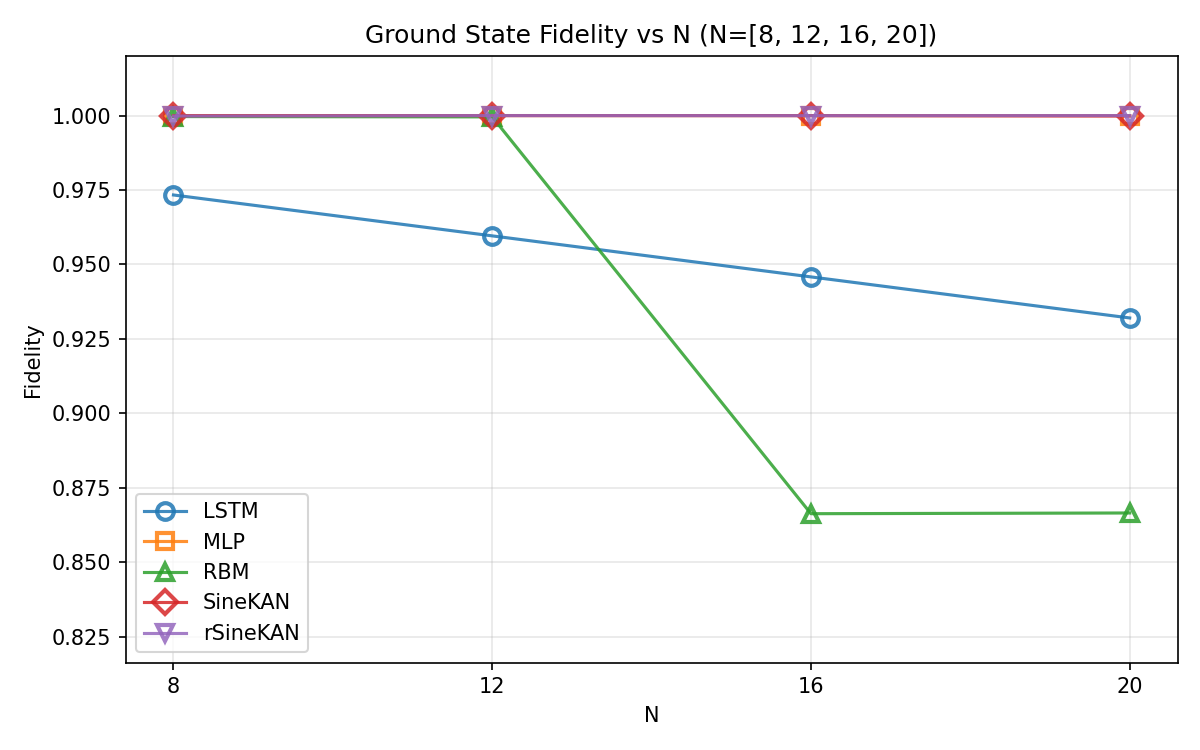}
\caption{Ground-state fidelity obtained with different positive-amplitude NQS architectures as a function of chain length $L$.}
\label{fig:pxp_fidelity}
\end{figure}

Figure~\ref{fig:pxp_fidelity} shows the ground-state fidelities of the different architectures for $L\in\{8,12,16,20\}$. MLP, SineKAN, and rSineKAN remain close to unity over the whole range. The LSTM fidelity decreases steadily with $L$. The RBM fidelity is close to unity at $L=8$ and $12$ but drops abruptly at $L=16$ and stays at the similar value at $L=20$. 

\subsection{Extended range with DMRG reference}

To examine larger systems, we extend the energy error and variance comparison to $L\in\{32,48,64\}$. The reference energy is the two-site DMRG value. For these DMRG calculations the bond dimension required never exceeds about $24$, well below the cutoff $\chi_{\max}=150$, consistent with the area-law entanglement expected of the PXP ground state. Because DMRG yields a matrix-product-state representation rather than an explicit state vector, the fidelity $F$ and the correlation function $C(r)$ are not evaluated. 

\begin{figure}[h]
\centering
\includegraphics[width=\textwidth]{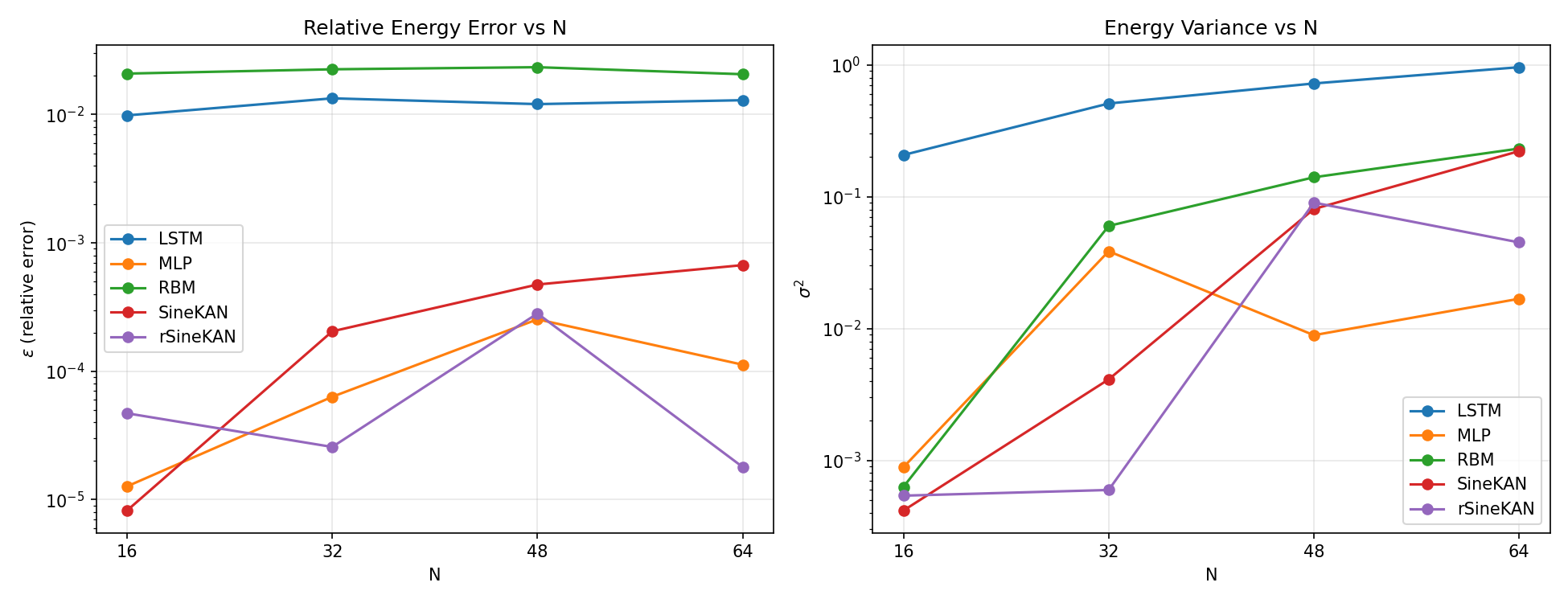}
\caption{Relative energy error $\epsilon$ (left) and energy variance $\sigma^2$ (right) for five positive-amplitude NQS architectures (LSTM, MLP, RBM, SineKAN, and rSineKAN) as functions of chain length $L\in\{16,32,48,64\}$, on a logarithmic scale. The reference energy is obtained from exact diagonalization for $L=16$ and from DMRG for $L>20$.}
\label{fig:pxp_error_variance}
\end{figure}

At $L=64$, LSTM, MLP, SineKAN, and rSineKAN have comparable numbers of parameters (LSTM: 83,480, MLP: 82,689, SineKAN: 82,929, rSineKAN: 82,929), while RBM has substantially more (532,544), since its number of weights grows quadratically with $L$ at fixed $\alpha$. Figure~\ref{fig:pxp_error_variance} shows $\epsilon$ and $\sigma^2$ for all five architectures.

The relative errors of RBM and LSTM stay roughly constant at the $10^{-2}$ level over the whole range. MLP, SineKAN, and rSineKAN have errors between $10^{-5}$ and $10^{-3}$, about two to three orders of magnitude smaller. The error of SineKAN grows steadily with $L$, from about $10^{-5}$ at $L=16$ to about $7\times10^{-4}$ at $L=64$. The errors of MLP and rSineKAN are non-monotonic: both peak at $L=48$, at about $3\times10^{-4}$, and decrease again at $L=64$, where rSineKAN has the smallest error among all five architectures.

The energy variance increases with $L$ for most architectures. For RBM, LSTM, and SineKAN it grows steadily by two to three orders of magnitude, and at $L=64$ the variances of RBM and SineKAN are comparable, while LSTM has the largest. The variance of rSineKAN stays near $10^{-3}$ up to $L=32$, rises sharply at $L=48$, and decreases slightly at $L=64$. The variance of MLP is non-monotonic and is the smallest of the five at $L=48$ and $L=64$.

\section{Conclusion}

The chiral symmetry of the PXP Hamiltonian fixes the sign structure of its ground state exactly: the sign of each amplitude is given by the parity of the total up-spin number. The associated parity transformation maps the PXP Hamiltonian to its negative, which is stoquastic and, since its configuration graph is connected, has a unique strictly positive ground state by the Perron--Frobenius theorem. Transforming back gives the exact sign rule of the original ground state, whose energy is identical to that of the transformed Hamiltonian.

The mapping also has methodological implications for variational Monte Carlo beyond computational convenience. In the original basis the ground state carries nontrivial relative signs, whereas after the parity transformation the Hamiltonian is stoquastic and the target state is strictly positive. The sign structure therefore does not have to be learned or encoded in the ansatz: it is absorbed into the Hamiltonian, so that a real, strictly positive variational state can be applied directly to the transformed Hamiltonian without any architecture-specific modification. We tested this with five neural quantum state architectures, SineKAN, reflection-symmetrized SineKAN, RBM, MLP, and LSTM, for different chain lengths. All five were trained in the same way on the transformed Hamiltonian. MLP, SineKAN, and reflection-symmetrized SineKAN reached high accuracy, while RBM and LSTM were less accurate. It would be interesting to explore whether similar constructions apply to other constrained systems.

\section*{Acknowledgments}
J.\ R.\ Park thanks the Korea Institute for Advanced Study for its associate member program. This work was supported by a research grant (K2536041) from the College of Public Policy at Korea University.
This research used resources of the National Energy Research Scientific Computing Center (NERSC), a Department of Energy Office of Science User Facility under Contract No. DE-AC02-05CH11231 using NERSC award DDR-ERCAP0037323.

\clearpage

\appendix

\section{Hyperparameters and computational settings}
\label{app:hyperparams}

\paragraph{Architectures and learning rates.}
Table~\ref{tab:hyperparams_small} lists the learning rate and architecture of each model for $L\in\{8,12,16,20\}$, used for Table~\ref{tab:pxp_summary} and Figures~\ref{fig:pxp_corr} and~\ref{fig:pxp_fidelity}. Table~\ref{tab:hyperparams_large} lists the corresponding settings for $L\in\{16,32,48,64\}$, used for Figure~\ref{fig:pxp_error_variance}. The RBM learning rate differs between the two sets ($10^{-3}$ and $10^{-5}$), whereas the other models use the same learning rate in both. For SineKAN and rSineKAN, the grid size at $L=48$ is set to $8$ instead of $7$, because $7=\mathrm{round}(\sqrt{48})$ and a grid size equal to $\mathrm{round}(\sqrt{L})$ was found to degrade the performance of SineKAN-based architectures~\cite{shamim2025probing}.

\begin{table}[h]
\centering
\caption{Hyperparameters used for $L\in\{8,12,16,20\}$.}
\label{tab:hyperparams_small}
\begin{tabular}{lcc}
\toprule
Model & Learning rate & Architecture \\
\midrule
RBM      & $1\times10^{-3}$ & $\alpha=128$ \\
SineKAN  & $1\times10^{-4}$ & grid size $=4$, hidden layers $=(60,60,1)$ \\
rSineKAN & $1\times10^{-4}$ & grid size $=4$, hidden layers $=(60,60,1)$ \\
MLP      & $1\times10^{-4}$ & hidden layers $=(140,140,1)$ \\
LSTM     & $1\times10^{-3}$ & features $=32$, layers $=4$ \\
\bottomrule
\end{tabular}
\end{table}

\begin{table}[h]
\centering
\caption{Hyperparameters used for $L\in\{16,32,48,64\}$.}
\label{tab:hyperparams_large}
\begin{tabular}{lcc}
\toprule
Model & Learning rate & Architecture \\
\midrule
RBM      & $1\times10^{-5}$ & $\alpha=128$ \\
SineKAN  & $1\times10^{-4}$ & grid size $=7$ ($8$ at $L=48$), hidden layers $=(80,80,1)$ \\
rSineKAN & $1\times10^{-4}$ & grid size $=7$ ($8$ at $L=48$), hidden layers $=(80,80,1)$ \\
MLP      & $1\times10^{-4}$ & hidden layers $=(256,256,1)$ \\
LSTM     & $1\times10^{-3}$ & features $=64$, layers $=4$ \\
\bottomrule
\end{tabular}
\end{table}

\paragraph{Optimization and sampling.}
All models are optimized with Adam at a fixed learning rate, without a learning-rate schedule, using the variational Monte Carlo implementation of NetKet~3.21. Each iteration uses $n_{\rm samples}=1024$ samples. RBM, MLP, SineKAN, and rSineKAN draw samples with MetropolisLocal using $1024$ parallel chains, and the LSTM uses autoregressive direct sampling. The number of optimization steps is $n_{\rm iter}=25000$ for all cases. The random seed is fixed for all runs, and each configuration is trained once.

\paragraph{Reference calculations.}
For $L\le20$ the reference ground state is obtained by Lanczos exact diagonalization of $\Hrot=-H$. For $L>20$ the reference energy is obtained by two-site DMRG (TeNPy) applied to the PXP Hamiltonian $H$, which gives the same ground-state energy as $\Hrot$ by Eq.~\eqref{eq:ground_relation}. We use a maximum bond dimension $\chi_{\max}=150$, a singular-value cutoff of $10^{-12}$, and an energy convergence tolerance of $10^{-12}$; the bond dimension actually reached does not exceed about $24$.

\clearpage

\bibliography{references_nqs}

\end{document}